\documentclass[fleqn,10pt,twocolumn]{wlscirep}
\usepackage{amssymb}
\usepackage{epsfig}
\usepackage{graphicx}
\usepackage{amsmath}
\usepackage{array,color}

\begin{document}

\title{Ground states of a Bose-Einstein Condensate in a one-dimensional laser-assisted optical lattice}
\author[1,*]{Qing Sun}
\author[1]{Jie Hu}
\author[2]{Lin Wen}
\author[3]{W.-M. Liu}
\author[4]{G. Juzeli\={u}nas}
\author[1,*]{An-Chun Ji}
\affil[1]{Department of Physics, Capital Normal University, Beijing 100048, China}
\affil[2]{College of Physics and Electronic Engineering, Chongqing Normal University, Chongqing, 401331, China}
\affil[3]{Beijing National Laboratory for Condensed Matter Physics, Institute of Physics, Chinese Academy of Sciences, Beijing 100190, China}
\affil[4]{Institute of Theoretical Physics and Astronomy, Vilnius University, Saul\.{e}tekio Ave. 3, LT-10222 Vilnius, Lithuania}

\affil[*]{sunqing@cnu.edu.cn,andrewjee@sina.com}

\begin{abstract}
We study the ground-state behavior of a Bose-Einstein Condensate (BEC) in a Raman-laser-assisted one-dimensional (1D) optical lattice potential forming a multilayer system. We find that, such system can be described by an effective model with spin-orbit coupling (SOC) of pseudospin $(N-1)/2$, where $N$ is the number of layers. Due to the intricate interplay between atomic interactions, SOC and laser-assisted tunnelings, the ground-state phase diagrams generally consist of three phases -- a stripe, a plane wave and a normal phase with zero-momentum, touching at a quantum tricritical point. More important, even though the single-particle states only minimize at zero-momentum for odd $N$, the many-body ground states may still develop finite momenta. The underlying mechanisms are elucidated. Our results provide an alternative way to realize an effective spin-orbit coupling of Bose gas with the Raman-laser-assisted optical lattice, and would also be beneficial to the studies on SOC effects in spinor Bose systems with large spin.
\end{abstract}







\flushbottom
\maketitle

\thispagestyle{empty}

\noindent 

\section*{Introduction}
The realization of Raman-induced artificial gauge fields in ultracold atomic gases \cite{Lin,ZhangJY,WangPJ,Cheuk,Qu,Huang,Wu} provides a well-controllable way to investigate many fundamental phenomena induced by SOC \cite{Dalibard,Goldman,Zhou,Zhai}. Among these studies, the spin-orbit (SO) coupled Bose gases, which have no counterpart in conventional solid materials, are of particular interests in cold atom community. An important consequence brought by SOC is the degeneracy in the single-particle ground states, which play a centre role in determining the many-body ground states of BECs. Many new phases as well as phase transitions are predicted to appear in diverse Bose systems with different types of SOC \cite{Wang,Li,Martone,Ho,Xu1,Wen,Zhang,Sinha,Hu,Wilson,Kawakami,Chen,Sun}. For example, the stripe and plane wave phases \cite{Wang,Li,Martone,Ho,Xu1,Wen,Zhang}, half-vortex (meron) ground states \cite{Sinha,Hu,Wilson,Kawakami,Chen}, and fractional skyrmion lattices \cite{Sun,Su} may emerge in SO coupled BECs.
 
Despite of different proposals to generate SOC in ultracold atoms \cite{Ruseckas,Stanescu,Jacob,Juzeliunas,Juzeliunas1,Campbell,Xu2,Xu3,Anderson,Anderson1,Liu}, so far  for Bose gases, the artificial SOC has been realized only in one dimension \cite{Lin,ZhangJY,Qu}  or in 2D lattices \cite{Wu}. Recently, the technique of laser-assisted tunneling \cite{Jaksch,Mueller,Gerbier} is developed to produce strong magnetic fields in optical lattices \cite{Aidelsburger1,Aidelsburger2,Miyake,Atala}. Such method provides a powerful and delicate way to manipulate atoms in lattice potential. Stimulated by these developments, some authors \cite{Sun,Su} have proposed an alternative and realistic way to realize an effective 2D SOC in bilayer Bose systems based on the laser-assisted tunneling. In such schemes, the prerequisite ``internal" states to fabricate SOC are essentially replaced by the Raman-assisted ``external" motional states in each layer, providing a new system to investigate the SO coupled BECs.

Motivated by the above advances, in this paper, we consider a gas of ultracold scalar bosons subjected to a Raman-assisted 1D optical lattice potential forming a multilayer system. Within the lowest band of the lattice, the system can be mapped to an effective model with SOC, where $N$ different layers play a role of a pseudospin $(N-1)/2$ coupled to the intralayer motion via the laser-assisted tunneling of atoms between the layers. This scheme can avoid the using of near resonant light beams which cause heating in previous experiments \cite{Lin}, and can be applied to a wide range of atom species including fermions. Recently, a related scheme has been experimentally implemented to realize the effective SOC with double well potential formed by an optical superlattice \cite{Ketterle}. In such a scheme the double layer is in a direction of the atomic motion. On the other hand, we suggest to use a lattice with $N$ sites in a direction perpendicular to the atomic motion. This resembles bosonic ladders \cite{Atala}, but the atoms now undergo a planar rather than a one-dimensional motion.

We determine the ground states of the system in the presence of atomic interactions. Note that, the dynamics of a SO coupled BEC in a weakly tilted optical lattice has been studied \cite{Larson}, where the correlated Bloch oscillations with spin Hall effect are revealed. Here, the one-dimensional optical lattice potential is sufficiently tilted and unlike the typical atom-atom interactions in conventional spinor BEC \cite{Ueda}, the special type of interactions from on-site repulsions in our system is quite different in the psuedospin representation, and can give rise to peculiar $N$-dependent phase diagrams with different behaviors: (1) For even $N$, by tuning the tunneling strength $J$, the single-particle ground state may change from a single minimal with zero-momentum to double minima with finite momentum, with the corresponding many-body ground states evolving from a normal phase to a robust stripe phase. (2) For odd $N$, the single-particle states only minimize at zero-momentum. However, when the interaction strength is increased, a stripe/plane wave phase with finite momentum can still emerge in the ground states. Such unique features reflect the competition and compromise between Raman-assisted tunneling and atomic interactions in this system.

\section*{Results}

\subsection*{The model}
\begin{figure}[t]
\centering
\includegraphics[width=0.95\linewidth]{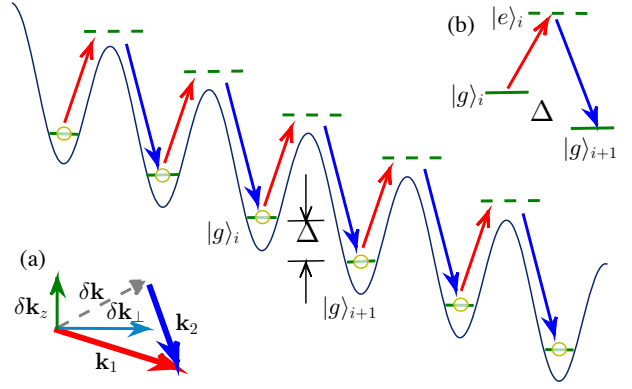}
\caption{Color online. Schematic diagram of the Raman-laser-assisted optical lattice. The lattice potential is deep and tilted enough so that the tight-binding approximation can be applied but the direct tunnelings between adjacent sites can be neglected. Two Raman lasers (labeled as red and blue arrows) couple the internal electron ground state $|g\rangle_i$ to an excited state $|e\rangle_i$ and would induce an interlayer transition (see the context for detail). (a). Energy levels with $\Delta$ the energy difference between adjacent sites. (b). Momentum relations. $\mathbf{k}_{1,2}$ are the incident momenta of two Raman lasers, $\delta\mathbf{k}=\mathbf{k}_1-\mathbf{k}_2=\delta\mathbf{k}_z+\delta\mathbf{k}_\perp$. }
\label{scheme}
\end{figure}

We consider a three-dimensional ultra-cold Bose gas (e.g. $^{87}$Rb) loaded into a one-dimensional optical lattice potential. Such potential are tight enough that the atoms only occupy the lowest energy band of the lattice potential (along $z$-axis), but move freely in the traverse $xy$-plane, forming a stacked-disk configuration. Furthermore, we apply a linear gradient potential in $z$-direction to tilt the lattice, as depicted in Fig. (\ref{scheme}). Such global tilt can be achieved by implementing a frequency shift between the lasers for the creating of the lattice potential \cite{Dahan,Morsch}, or by tilting the lattice along the direction of the gravitational field \cite{Fattori,Gustavsson}. The single-particle Hamiltonian of this system reads:
\begin{eqnarray}
H_0&=&\int d^3\mathbf{r}\Psi^\dag(\mathbf{r}) \{\frac{\mathbf{P}^2}{2m}+U(z)\}\Psi(\mathbf{r})
\label{H1}
\end{eqnarray}
with
\begin{eqnarray*} 
U(z)&=&U_0\cos^2(k_oz)+V(z)-Fz,
\end{eqnarray*}
where $\Psi(\mathbf{r})$ annihilates a boson at position $\mathbf{r}$. $U_0$ and $F$ are the strengths of optical and linear gradient potential respectively, and $V(z)=\frac{1}{2}\omega^2_zz^2$ is a weak harmonic potential along z-axis with $\omega_z$ the trapping frequency. $\mathbf{P}$ and $m$ are the momentum and mass of atom, and $k_o$ is the wave-vector of laser to generate the lattice potential. 

When the tilting is not too large, the atoms can still move in the lowest state of each well, forming the energy band of the lattice potential. We can expand the field operator $\Psi(\mathbf{r})=\sum^N_i\phi_i(x,y)w(z-z^c_i)$ with the localized wannier function $w(z-z^c_i)$ of the $i$th lattice, where $N$ is the lattice number. The Hamiltonian. (\ref{H1}) then can be rewritten as:
\begin{eqnarray}
H^\prime_0=\int d^2\mathbf{r}\sum^N_i\phi^\dag_i(x,y) \{\frac{\mathbf{P}^2_{\perp}}{2m}+\epsilon_i\}\phi_i(x,y).
\label{H2}
\end{eqnarray}
Here, $\epsilon_i=\int dzw^*(z-z^c_i)\{\frac{P^2_z}{2m}+U_0\cos^2(k_oz)+V(z)-Fz\}w(z-z^c_i)$ is the on-site energy and $\mathbf{P}^2_{\perp}=\mathbf{P}^2_x+\mathbf{P}^2_y$.
When neglecting the small $V(z)$, the energy differences between adjacent sites are mainly caused by the linear-tilted part, which is given by  $\Delta\equiv\Delta_{ij}=|\epsilon_i-\epsilon_{j}|$. Generally, there should also be a tunneling matrix element $\mathcal{K}\equiv\mathcal{K}_{ij}=\int dzw^*(z-z^c_i)\{\frac{P^2_z}{2m}+U_0\cos^2(k_oz)-Fz\}w(z-z^c_j)$ between adjacent sites $<ij>$. However, for a sufficiently tilted lattice potential, the inter-site tunneling is much smaller than the energy mismatch between two sites, i.e. $\mathcal{K}\ll\Delta$. For example,  $\mathcal{K}\approx 2.5\times 10^{-3}E_R$ for $\Delta\approx 8E_R$ with $E_R$ the recoiling energy of the lattice \cite{Gerbier}. As a result, the direct tunneling is inhibited and hence can be neglected. To restore the atomic hopping between adjacent wells, we resort to the newly developed laser-assisted process \cite{Aidelsburger1,Aidelsburger2,Miyake,Atala}. 

To this end, we implement two Raman lasers with wave-vector $\mathbf{k}_i$ and frequency $\omega_i$ ($i=1,2$), which couple to the atomic internal state via a two-photon transition. This gives rise to an time-dependent scalar potential $V_K=V_0[e^{i(\delta\mathbf{k}\cdot\mathbf{r}-\delta\omega t)}+e^{-i(\delta\mathbf{k}\cdot\mathbf{r}-\delta\omega t)}]$, where $V_0$ is controlled by the Raman beam intensities. $\delta\mathbf{k}=\mathbf{k}_1-\mathbf{k}_2$ and $\delta\omega=\omega_1-\omega_2$ denotes the wave vector and frequency differences of the Raman lasers. Then along the z-direction, one has an additional overlap integral $\mathcal{K}^\prime_{ij}=V_0\int dzw^*(z-z^c_i)w(z-z^c_j)[e^{i(\delta k_z z+\delta k_x x+\delta k_y y-\delta\omega t)}+e^{-i(\delta k_z z+\delta k_x x+\delta k_y y-\delta\omega t)}]$, i.e. $\tilde{H}^\prime_0=H^\prime_0+K^\prime=H^\prime_0+\sum_{ij}(\mathcal{K}^\prime_{<ij>}+h.c.)$, where we have assumed $\delta\omega\sim\Delta$ with the dominant contribution in $K^\prime$ is from the overlap between adjacent sites $<ij>$, while other processes are far off-resonance and can be neglected. Then by introducing $N$-component spinor $\Phi=(\begin{array}{cccc}
\phi_1 \phi_2 \cdots \phi_N
\end{array})^T$, and applying a unitary transformation $U=e^{-i\hat{S}}$, where 
\begin{eqnarray}
\hat{S}=\int d^2\mathbf{r}(\hbar\delta\mathbf{k}_\perp\cdot\mathbf{r}_\perp-\delta\omega t)\Phi^\dag\hat{F}_z\Phi
\end{eqnarray}
and $\delta\mathbf{k}_\perp\cdot\mathbf{r}_\perp=\delta k_x x+\delta k_y y$, the Hamiltonian is transformed as $\tilde{H}^\prime_0\rightarrow U\tilde{H}^\prime_0U^\dag+(i\partial_tU)U^\dag$ (The definition of $F_z$ is referred to Eq. 6). To write it explicitly, we have $UH^\prime_0U^\dag+(i\partial_tU)U^\dag=\int d^2\mathbf{r}\Phi^\dag[\frac{(\mathbf{P}_\perp+\hbar\delta\mathbf{k}_\perp\hat{F}_z)^2}{2m}+(\Delta-\delta\omega)\hat{F}_z]\Phi$, and $UK^\prime U^\dag=\sum\limits_{<ij>}[(J+J^*e^{-2i(\hbar\delta\mathbf{k}_\perp\cdot\mathbf{r}_\perp-\delta\omega t))})\phi^\dag_i\phi_j+h.c.]$ with $J\equiv \mathcal{K}^\prime_0=V_0\int dzw^*(z-z^c_i)w(z-z^c_j)e^{i\delta k_z z}$ being the laser-assisted tunneling strength. Notice that, it is the factor $e^{i\delta k_z z}$ due to the momentum transfer along the z-axis making the overlap integral $J$ nonzero. Under the rotating-wave approximation, one can drop the counter-rotating terms and arrive at the following effective single-particle Hamiltonian
\begin{eqnarray}
H^{\rm eff}_0=&&\int d^2\mathbf{r}\Phi^\dag[\frac{(\mathbf{P}_\perp+\hbar\delta\mathbf{k}_\perp\hat{F}_z)^2}{2m}+J\hat{\mathcal{M}}\nonumber\\
&&+\delta_\epsilon\hat{F}_z+\delta_z\hat{F}^2_z]\Phi.
\label{effective}
\end{eqnarray}
Here, 
\begin{eqnarray}
\hat{\mathcal{M}}=\left(\begin{array}{cccc}
0 & 1 & \cdots & 0\\
1 & 0 & \cdots & 0\\
0 & 1 & \cdots & 1\\
0 & 0 & \cdots & 0 
\end{array}\right),
\end{eqnarray} 
and 
\begin{eqnarray}
\hat{F}_z=\left(\begin{array}{cccc}
\frac{N-1}{2} & 0 & \cdots & 0\\
0 & \frac{N-3}{2} & \cdots & 0\\
0 & 0 & \cdots & 0\\
0 & 0 & \cdots & -\frac{N-1}{2} 
\end{array}\right)
\end{eqnarray} is the $z$- component of angular momentum matrix with angular momentum $L=\frac{N-1}{2}$, $\delta_\epsilon=\Delta-\delta\omega$ is the two-photon detuning. Furthermore, we have also included an additional harmonic trap $V(z)$, which can be termed as an effective quadratic Zeeman energy $\delta_z=\frac{1}{2}\omega^2_za^2$ ($a$ is the lattice spacing) by adjusting the trapping center. Due to $[\mathcal{M}, F_z]\equiv\mathcal{M}F_z-F_z\mathcal{M}\neq 0$, Eq. (4) effectively describes a system with nontrivial spin-orbit coupling, which reduces to a familiar form with equal Rashba and Dresselhaus contributions \cite{Li,Martone,Ho} for $N=2$ and $N=3$. The major difference is that the internal spin states now are played by atoms in different wells and the total pseudospin can be varied via the lattice number $N$. 

Before proceeding, we should mention that there are generally Bloch oscillations for atoms in a tilted optical lattice potential \cite{Larson,Dahan,Morsch,Fattori,Gustavsson}. For sufficiently tilted lattice case as discussed above, the energy difference $\Delta$ between adjacent sites is much larger than the suppressed inter-site tunneling $\mathcal{K}$. In this regime, the Bloch oscillation becomes a rapid shivering motion with frequency $\Delta/h$ and amplitude $\mathcal{K}/\hbar$. When the two-photon Raman transition is introduced, $\Delta$ reduces to a two-photon detuning $\delta_\epsilon$, and a considerable effective tunneling $J$ is induced. In this case, it was shown that the coherent Bloch oscillations of frequency $\delta_\epsilon/h$ would appear in the form of a periodic breathing dynamics \cite{Alberti} or a periodic center of mass motion \cite{Haller}. In the following, we mainly concentrate on the ground-state behaviors for the resonant case with $\delta_\epsilon=0$.

\begin{figure}[t]
\centering
\includegraphics[width=\linewidth]{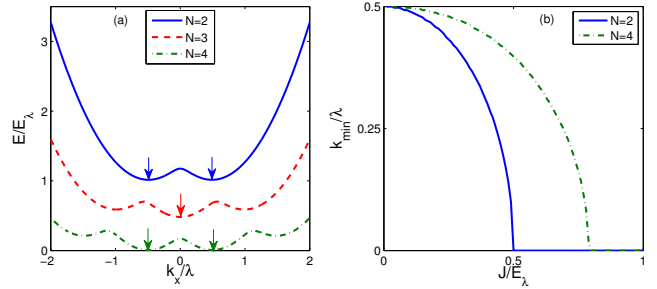}
\caption{Color online. (a). The lowest single-particle energy branch of $N=2$ (blue solid), 3 (red dashed), 4 (green dash-dotted) for $J/E_\lambda=0.1$. The curves have been shifted constantly for explicitness. Arrows label the energy minima. (b). Momentum evolution of the minimal states with the tunneling strength $J$ for $N=2$ (blue solid) and $N=4$ (green dash-dotted). In both figures (a) and (b) the relative strength of the quadratic Zeeman shift is taken to be $\delta_z/E_\lambda=0.1$.}
\label{single_particle}
\end{figure}

Without loss of generality, in the following we assume that the traverse momentum transfer $\delta\mathbf{k}_\perp$ is along the x-axis, i.e. $\delta\mathbf{k}_\perp=\lambda \hat{\mathbf{e}}_x$ with $\lambda=|\delta\mathbf{k}_\perp|$, and set $m=\hbar=1$ and energy unit $E_\lambda=\lambda^2/2$ throughout the paper. 

\subsection*{Single particle spectrum}

We first discuss the single particle states of this system. In the absence of effective Zeeman fields, Hamiltonian. (\ref{effective}) bears the time-reversal symmetry (TRS) with $H(-\mathbf{k},-\hat{F}_z)=H(\mathbf{k},\hat{F}_z)$, resulting in a symmetrical single-particle energy spectrum $E(\mathbf{k})=E(-\mathbf{k})$. Due to the laser-assisted tunneling $J$, the atomic states in different wells get mixed, and hence the degeneracy of psuedospin components is lifted with $N$ energy branches. We are interested in the lowest branch, which is responsible for the determining of bosonic ground states. 

In general, the single-particle ground-state manifold can be classified into two categories: for even $N$, there may exist a two-fold degeneracy; while for odd $N$, there is only one state in the ground subspace. As shown in Fig. (2a), we plot the lowest energy spectrum by diagonalizing Hamiltonian. (\ref{effective}) for different $N$ ($=2,3,4$) with $J/E_\lambda=0.1$. We can see that, for odd $N$ ($=3$), there is only one minimum state at $k=0$. On the other hand, for even $N$ ($=2,4$), double minima at $\pm k_{\rm min}$ can be identified. Here, we have also included a weak harmonic trap. As the quadratic Zeeman term does not break the TRS, it would just modify the ground state energy for small $\delta_z$ without destroying the double degeneracy. On the other hand, a nonzero linear term $\delta_\epsilon\neq 0$ would break the TRS and lead to asymmetric energy spectra. As a result, such possible degeneracy is lifted, with only one state left in the ground-state manifold.

In Fig. (2b), we plot the momentum evolution of minimal states as a function of $J$ for even $N$ ($=2,4$). One can find that, when $J$ surpasses a critical $J_c$, $k_{\rm min}$ would converge to $0$, indicating a tunneling induced transition would happen in the single-particle ground state. As we will see below, above different behavior of the single-particle states would have dramatic effects on the many-body ground states when the atomic interactions are included. 

\subsection*{Ground state phase diagram}
We now turn to investigate the many-body ground states of this system in the presence of atom-atom interactions. Considering a short-ranged case, the interactions for atoms situated in the same well are much stronger than that in different wells. Then, one can neglect the contribution from the latter and write the Hamiltonian for interacting atoms as
\begin{eqnarray}
H_{\rm int}=g_{\rm 2D}\int d^2\mathbf{r}\sum^N_in^2_i,
\end{eqnarray}
where $n_i=\Phi^\dagger_i\Phi_i$ denotes the atomic density in $i$th layer and $g_{\rm 2D}=\frac{g_0}{2}\int dz|w(z-z^c_i)|^4$ with $g_0$ the contact interaction strength. Notice that, the interactions here only happen in each pseudospin component, which keeps invariant under the unitary transformation $U$, and would play an important role in determining the ground-state configurations. 

\begin{figure}[t]
\centering
\includegraphics[width=\linewidth]{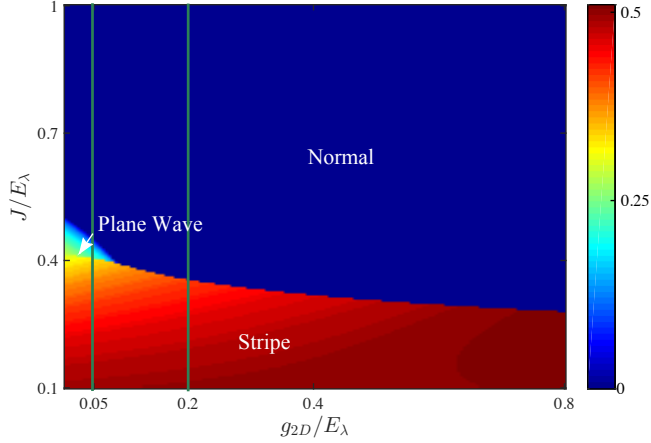}
\caption{Color online. Phase diagram in $g_{\rm 2D}-J$ plane for $N=2$, consist of three phases: Stripe, Plane Wave and Normal phases (see the context for detail), touching at a tricritical point. The color bar denotes the magnitude of ground state momentum. The green solid lines label the parameters we used in Fig.(\ref{even_property}).}
\label{even}
\end{figure}

In the following we will discuss independently the ``even" and ``odd" N cases, which exhibit different behavior in the single-particle spectra and the many-body ground states.

\subsubsection*{Even $N$}

In this case, we find that for giving trapping potential $\delta_z$, the phase diagrams in $g_{2D}-J$ plane for different $N$ have similar structures, and three different phases may appear: (I) ``Stripe" phase, where the wave-function is a superposition of two plane waves with opposite momenta $\pm k_m$ ($k_m\neq 0$) and $a_+=a_-=1/\sqrt{2}$; (II) ``Plane Wave" phase, where only one plane wave component with finite momentum $k_m$ contributes to the ground state; (III) a ``Normal" phase with bosons condensed in the zero-momentum state of $k=0$. 

To be more specific and without loss of generality, we choose the simplest $N=2$ for illustrations. In Fig. (3), we give the ground-state phase diagram in the $g_{\rm 2D}-J$ plane for $N=2$ by numerically minimizing the energy $E_{\rm G}$. Generally, due to the interplay between atomic tunneling and atom-atom interactions, above three phases may compete with each other and survive in three distinct regimes (labeled by colors), touching at a tricritical point.
 
In the dilute limit ($g_{\rm 2D}/E_\lambda\ll1$), above a critical tunneling strength, i.e. $J>J_{c1}\simeq 0.5E_\lambda$, the system is in the zero-momentum Normal phase. While for $J<J_{c2}\simeq 0.41E_\lambda$, a Stripe phase is favored. Between them ($J_{c1}<J<J_{c2}$), a Plane Wave phase is expected to have lower energy. The regime of such Plane Wave phase gets diminished with increasing of interaction $g_{\rm 2D}$, and finally disappears at a tricritical point around $(J/E_\lambda, g_{\rm 2D}/E_\lambda)\simeq(0.38, 0.11)$, where three phases merge. Beyond the tricritical point, only Normal to Stripe phase transition survives (see Fig. 4(b,d)). These features essentially reflect the competitions between kinetic and interaction energies of these states. In the weak interaction regime, the kinetic energy is dominant, and the system is always in a Normal phase when the single particle spectrum has only one minimum at $k=0$. On the double minima side, the kinetic energies of Stripe and Plane Wave phases for the same $k_m$ are degenerate, and would be further lifted by the atomic interactions. 

\begin{figure}[t]
\centering
\includegraphics[width=\linewidth]{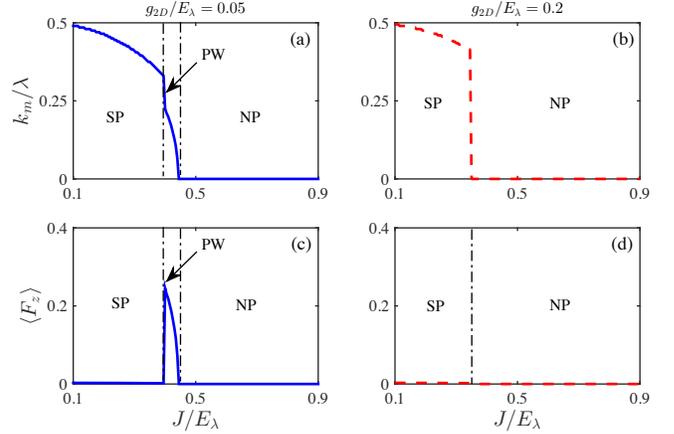}
\caption{Color online. The ground state momentum $k_m/\lambda$ (a,b) and interlayer polarization $\langle F_z\rangle$ (c,d) as functions of tunneling $J/E_\lambda$ for given interactions $g_{\rm 2D}/E_\lambda=0.05$ (left panel) and $g_{\rm 2D}/E_\lambda=0.2$ (right panel) at $N=2$. SP, PW and NP denote Stripe, Plane Wave and Normal phases, respectively.}
\label{even_property}
\end{figure}

In Fig. (\ref{even_property}), we plot the ground-state momenta $k_m$ and the interlayer polarization $\langle F_z\rangle$ as functions of tunneling $J$ for two typical interaction strength. One can see that, the Plane Wave has homogeneous intralayer density $\bar{n}_i$ but finite interlayer polarization $|\langle F_z\rangle|> 0$, while the Stripe phase has inhomogeneous density $n_i(r)$ with $\langle F_z\rangle=0$. On the other hand, since the atoms in the same layer repulse each other, the system tends to have both equal populations and homogeneous densities in each layer. Hence, close to the Normal phase, the Plane Wave phase with small $|\langle F_z\rangle|$ but homogeneous intralayer density is more favorable. While with decreasing of $J$, $|\langle F_z\rangle|$ becomes larger and larger, and the system transits into the Stripe phase. Moreover, the Plane Wave phase would be also suppressed by increasing of interactions and turn to a Normal phase continuously. Such two transitions finally meet at a quantum tricritical point.

Several remarks are on hand: first, we have taken $\delta_z/E_\lambda=0.01$ in numerical calculations, which gives no physical effects for $N=2$ and would modify the phase boundaries slightly for $N>2$. Second, compared to the effective model in \cite{Lin}, here the {\it external states} in different layers play the role of spin rather than the {\it internal states}. The corresponding effective spin-spin interaction takes the value $c_2/c_0=(g_\uparrow-g_{\uparrow\downarrow})/(g_\uparrow+g_{\uparrow\downarrow})=1$ with $g_{\uparrow\downarrow}=0$, which is much larger than in the previous case, where $c_2$ is very close to the degenerate point $c_2=0$ \cite{Lin,Sun}. This makes the Stripe phase in this system quite robust \cite{Li}. Third, for $N>2$, the phase diagrams are qualitatively unchanged around tricritical regime and similar analysis can be applied. 

\begin{figure}[t]
\centering
\includegraphics[width=\linewidth]{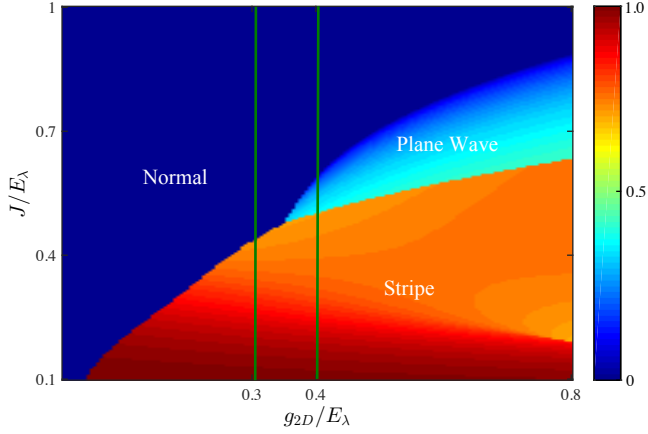}
\caption{Color online. Phase diagram in $g_{\rm 2D}-J$ plane for $N=3$. The color bar denotes the momentum magnitude of ground states. The green solid lines label the parameters we used in Fig.(\ref{odd_property}).}
\label{odd_N}
\end{figure}

\subsubsection*{Odd $N$}

When $N$ is odd, the situation changes a lot and the phase diagrams may exhibit different behaviors. To be specifc, in the following we take $N=3$ as an example to address this problem. Similar results can be found for $N>3$.

\begin{figure}[t]
\centering
\includegraphics[width=\linewidth]{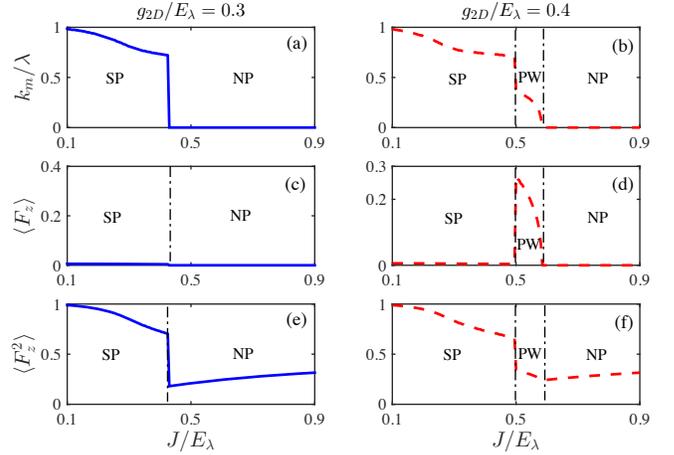}
\caption{Color online. The ground state momentum $k_m/\lambda$ (a,b), $\langle F_z\rangle$ (c,d) and $\langle F^2_z\rangle$ (e,f) as functions of tunneling $J/E_\lambda$ for given interactions $g_{\rm 2D}/E_\lambda=0.3$ (left panel) and $g_{\rm 2D}/E_\lambda=0.4$ (right panel). Here, $N=3$ and $\delta_z/E_\lambda=0.01$.}
\label{odd_property}
\end{figure}

In Fig. (\ref{odd_N}), we give the phase diagram for $N=3$. It is interesting to see that even though the single-particle spectrum is only minimized at the  $k=0$ state, the system can still be Plane Wave or Stripe phases carrying finite momenta in some regimes. In one hand, the zero-momentum Normal phase is predominant for small $g_{\rm 2D}$. In the other hand, the interaction energy would become significant with the increasing of interactions. As shown in Fig. (\ref{odd_property}e,f), the density of Normal phase $n_0=1-\langle F^2_z\rangle$ in center layer is relative large. And for sufficient large $g_{\rm 2D}$, an instability to Plane Wave/Stripe phases with more delocalized atomic distribution and finite $k_m$ (Fig. (\ref{odd_property}a,b)) would happen, where the increasing of kinetic energies is compensated by the decreasing of interaction energies.
Furthermore, similar to the even $N$ case, the Plane Wave phase with a finite $\langle F_z\rangle\neq0$ (Fig. (\ref{odd_property}c,d)) only survives for moderate tunneling strength $J$, between the Normal and Stripe phases, and ends at a tricritical point.

It is worthy to stress that, the emerging of Plane Wave/Stripe phases for odd $N$ is mainly driven by atom-atom interactions, in a sharp contrast to the even $N$ case, where the role is mainly played by atomic tunnelings. This may reflect the topological differences of single-particle ground-state manifolds between these two cases.

\section*{Discussions and conclusions}
We now discuss some experiment-related issues. First, our results are quite general and independent of specific atoms. Here, we take the $^{87}$Rb as an example. The simplest $N=2$ case can be achieved by a similar scheme as the bilayer configurations \cite{Sun}. For $N>2$, one can resort to a superlattice potential with more than two nonequivalent sites \cite{Gerbier} or a linear tilt potential \cite{Jaksch}. For a standing wave with wave-length $\lambda_s$ and depth $U_0\sim 15E_r$,  where $E_r=h^2/(2m\lambda^2_s)$ is the recoil energy, the trapping frequency in each well is about $\omega_o\sim\sqrt{4E_rU_0}$. If one choose $\Delta\sim E_r$, the Raman-assisted tunneling $J\sim \frac{J_zV_0}{\Delta}$ with the bare tunneling $J_z\ll\Delta$, can be tuned up to $J\sim 2\pi\times60$ Hz by varying $V_0$. Note that, $V_0\ll\Delta\ll\omega_o$ can be satisfied to ensure the validity of tight-binding and the lowest band approximations. To reach the scope of the phase diagram, one need $E_\lambda\sim J$. This can be done by arranging the opening angles of two Raman lasers with $|\delta\mathbf{k}_\perp|\sim\sqrt{2mJ/\hbar^2}$. For a typical harmonic trap with frequency $\omega_z\sim2\pi\times10$ Hz, $\delta_q\sim m\omega^2_z\lambda^2_s/8$ is much smaller than $E_\lambda$. In the case of $^{87}$Rb, $g_0\sim7.8\times 10^{-12}$Hz cm$^3$, and the corresponding $g_{\rm 2D}=\sqrt{2\pi}N_ag_0/2\xi_z$ with $\xi_z=\sqrt{\hbar/m\omega_o}$, is limited to a weakly interacting regime. 

Up to now, we have neglected the effects of effective Zeeman fields $\delta_\epsilon$ and $\delta_q$. For not too large $\delta_\epsilon$ and/or $\delta_q$, the phase boundaries would be modified quantitatively \cite{Wen} which are also confirmed in our case, while leaving the main results qualitatively unchanged. To detect these phases in experiments, one may implement the momentum-resolved time-of-flight measurements. The atom population in each well which characterizes $\langle F_z\rangle$ and $\langle F^2_z\rangle$, can be measured via ${\it in}$-${\it situ}$ absorption imaging. 

In conclusions, we have investigated the ground states and the associated phase diagrams of a BEC in a laser-assisted 1D optical lattice potential forming a multilayer system. The unique $N$-dependence of the single-particle spectra and the corresponding many-body ground-state configurations reflects the subtle competition between the effective SOC induced by laser-assisted interlayer tunneling and atom-atom interactions. Our results would have potential implications in searching new matter states in spin-orbit coupled Bose systems with a large spin. In future studies, one may consider the effects of interlayer long-range interactions, and the extensions to multicomponent BECs and Fermi gases.

\section*{Methods}

For weakly interacting Bose gases, the quantum fluctuations can be neglected safely. And one can adopt the variational method \cite{Li,Sun} to investigate the ground states of the system. In mean-field level, the variational Ansatz of the ground-state wave-function can be constructed as: 
\begin{eqnarray}
\Psi_{\rm G}=a_+\Phi_{k}e^{ikx}+a_-\Phi_{-k}e^{-ikx},
\end{eqnarray}
where $\Phi_{\pm k}e^{\pm ikx}$ are the eigenstates of the lowest energy branch with momentum $\pm k$, determined by Eq. (\ref{effective}). $a_+$ and $a_-$ are complex amplitudes with normalization condition $|a_+|^2+|a_-|^2=1$. In the dilute limit, one has $k=k_{\rm min}$ for double minima case and $k=0$ for single minimum. While in general, $k$ is dependent on the interactions \cite{Li}.  Minimizing the energy $E_{\rm G}=\langle\Psi_{\rm G}|H^{\rm eff}_0+H_{\rm int}|\Psi_{\rm G}\rangle$ with respect to variational parameters $a_+$, $a_-$ and $k$, one can obtain the ground-state phases as well as the phase diagrams.

\section*{Acknowledgement}
This work is supported by NSFC (Grants No. 11404225, No. 11474205, No. 21503138, No. 11504037 and No. 11434015), Foundation of Beijing/Chongqing Education Committees (Grants No. KM201510028005, KM201310028004, and No. KJ1500311), and CSTC under Grant No. cstc2015jcyjA50024. G. J. acknowledges a support by Lithuanian Research Council (Grant No. MIP- 086/2015).

\section*{Author contributions statement}
Qing Sun performed the theoretical as well as the numerical calculations. Qing Sun, Jie Hu, Lin Wen, W.-M. Liu, G. Juzeli\={u}nas and An-Chun Ji wrote and reviewed the manuscript.

\section*{Additional information}
\textbf{Competing financial interests:} The authors declare no competing financial interests.

\end{document}